\documentclass{ifacconf}

\usepackage{graphicx} 
\usepackage{natbib} 
\usepackage{mathtools}
\usepackage{amssymb}
\usepackage{array}
\usepackage{booktabs}
\usepackage{cuted, lipsum}
\usepackage{xcolor}

\usepackage{breqn}
\usepackage{tikz}
\usetikzlibrary{graphs,quotes,arrows.meta}

\begin{document}
\begin{frontmatter}

\title{Parallel-Connected Battery Current Imbalance Dynamics}

\author[First]{Andrew Weng} 
\author[First]{Sravan Pannala}
\author[First]{Jason B. Siegel}
\author[First]{Anna G. Stefanopoulou} 

\address[First]{
   University of Michigan, 
   Ann Arbor, MI 48109, USA \\ 
   (e-mail: asweng@umich.edu)
   }

\begin{abstract}
    
    In this work, we derive analytical expressions governing state-of-charge and current imbalance dynamics for two parallel-connected batteries. The model, based on equivalent circuits and an affine open circuit voltage relation, describes the evolution of state-of-charge and current imbalance over the course of a complete charge and discharge cycle. Using this framework, we identify the conditions under which an aged battery will experience a higher current magnitude and state-of-charge deviation towards the end of a charge or discharge cycle. This work enables a quantitative understanding of how mismatches in battery capacities and resistances influence imbalance dynamics in parallel-connected battery systems, helping to pave a path forward for battery degradation modeling in heterogeneous battery systems. 
    
\end{abstract}


\end{frontmatter}

\section{Introduction}

\begin{figure} 
\begin{center}
\includegraphics[width=6.0cm]{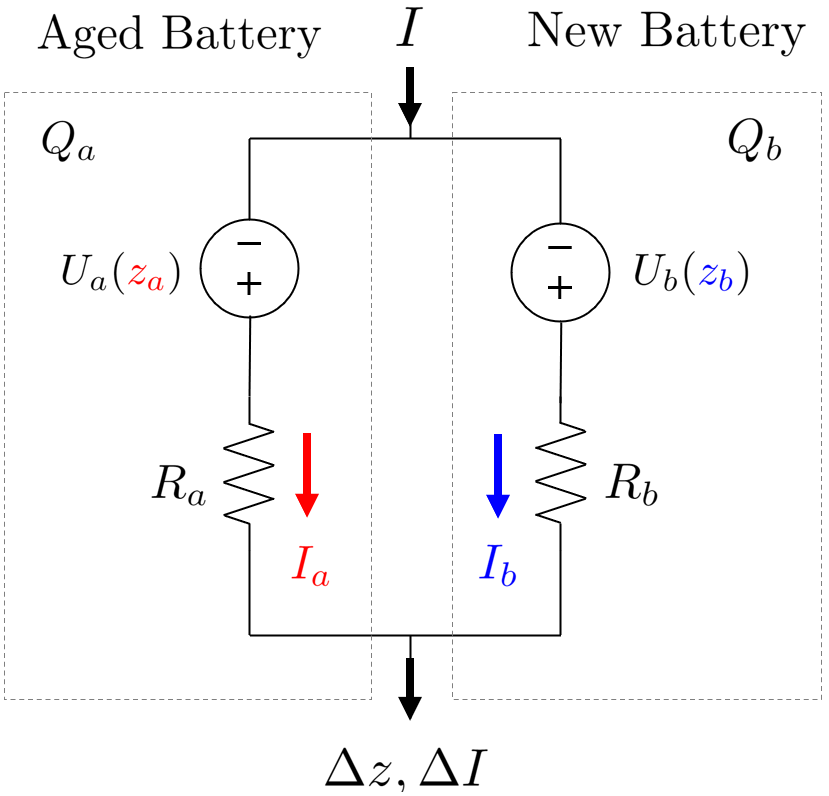} 
\caption{Two non-identical batteries are connected in parallel. When cycled together, how do mismatches in initial states of charge ($z$), capacities ($Q$) and resistances ($R$) affect the current and state of charge imbalance?}
\label{fig:circuit}
\end{center} 
\end{figure}

Battery degradation behavior is often understood in the context of single battery cells. Yet, under real applications, batteries are often connected in parallel to increase available system capacity and power. When non-identical batteries are connected and cycled in parallel, as shown in Figure \ref{fig:circuit}, complex current and state-of-charge (SOC) dynamics can arise due to mismatches in capacities and resistances of the individual batteries. Such dynamics can cause one battery to experience higher peak currents or to become more charged at the end of a charge or discharge, leading to more degradation relative to the other batteries. Several key questions arise: in a parallel-connected battery pack with mismatched battery capacities and resistances, do the degradation trajectories of individual batteries converge or diverge over life? Such questions are foundational to understanding battery management system requirements for aged battery packs. If non-identical batteries cause divergent aging behavior, for example, then efforts to manage second-life battery pack health may become complex and costly. If, however, the aging behavior is shown to be convergent, then heterogeneous battery packs may be able to `self-regulate', improving the feasibility of deploying such systems in the field. These questions are particularly salient for second-life applications in which aged batteries may be repurposed and reused alongside fresh batteries in a single pack \citep{Zhu2021}. 

The behavior of parallel-connected batteries has been explored empirically by \cite{Luca2021, Schindler2021a, Diao2019}, who independently demonstrated the potentially significant impact of mismatched resistances and capacities on the current imbalance at the end of a charge cycle. Model-based approaches have also been proposed for pack-level simulations and controls. \cite{Bruen2016} and \cite{Gong2015} developed equivalent-circuit based models enabling simulation of current dynamics in parallel-connected systems. Similar models have also recently been applied to develop SOC estimators \citep{Zhang2020g}, study the influence of cell-to-cell variances on the thermal design of battery packs \citep{Fill2021}, and simulate the aging behavior of grid battery systems \citep{Reiners2022}. However, despite the existing literature, fundamental analysis of such systems remain challenging due to a lack of first-principles descriptions of current-sharing behavior in parallel-connected batteries.

In this work, we focus on deriving analytical equations for current and SOC imbalance for two parallel-connected batteries and analyzing the system's dynamical properties. Our characterization provides an intuitive but quantitative understanding of how differences in battery cell capacities and resistances influence the growth rate and magnitude of current and SOC imbalance over the course of a complete charge and discharge cycle. Our analysis requires a linearization of the cell open circuit potential function, so we provide a discussion on model accuracy when compared to a non-linear case. We finally discuss the degradation circumstances that could lead to convergent or divergent degradation pathways for mismatched cells connected in parallel. This work is foundational to understanding how current and SOC imbalance can lead to different degradation outcomes in parallel-connected battery systems.

\section{Problem Formulation}

For this study, we start with the simplest possible representation of a parallel-connected battery system: two battery cells, each represented by an OCV-R model, connected in parallel as shown in Figure \ref{fig:circuit}. We define the following quantities for each cell $i \in \{a,b\}$:

\begin{itemize}
    \item $z_i(t) \in [0,1]$ : cell state of charge (SOC) at time $t$
    \item $U_i(z_i) \in \mathbb{R^+}$ : cell open-circuit voltage (OCV)
    \item $R_i \in \mathbb{R^+}$ : cell internal resistance
    \item $Q_i \in \mathbb{R^+}$ : cell capacity
    \item $I_i \in \mathbb{R^+}$ : cell branch current
\end{itemize}

$R_i$ and $Q_i$ are assumed to be constants. $U_i$ is generally a non-linear, monotonically increasing function. The system input is the applied current $I$ which is defined to be negative for charge and positive for discharge. The system output is the terminal voltage, $V_t$, which is identical for all batteries due to voltage conservation, and is given by

\begin{align}
    \label{eqn:kvl}
    V_t &= U_i(z_i) - I_iR_i
    .
\end{align}

Current conservation further requires that

\begin{equation}
    \label{eqn:kcl}
    I = \sum_i I_i
    .
\end{equation}

Applying (\ref{eqn:kvl}) and (\ref{eqn:kcl}) for two parallel-connected cells yields the following relationships for the branch currents and terminal voltage as a function of the input current $I$:

\begin{align}
    \label{eqn:vt}
    V_t &= \frac{R_bU_a(z_a) + R_aU_b(z_b) - R_aR_b I}{\mathcal{R}} \\
    \label{eqn:ia}
    I_a &= \frac{U_a(z_a) - U_b(z_b) + R_bI}{\mathcal{R}} \\
    \label{eqn:ib}
    I_b &= \frac{U_b(z_b) - U_a(z_a) + R_aI}{\mathcal{R}}
\end{align}

where

\begin{equation}
    \mathcal{R} \triangleq R_a + R_b.
\end{equation}

A key simplification we will make throughout the this work is to assume that the cells have identical and affine OCV functions which take the form

\begin{align} 
    \label{eqn:ocv-affine}
    U_i = U^0 + \alpha z_i
\end{align}

where $U^0, \alpha \in \mathbb{R}^+$. $U^0$ is the minimum voltage definition for the battery and $\alpha$ characterizes the slope of the OCV function. A discussion of the case of a non-linear OCV function is provided in Section \ref{sec:nonlin}. For this work, we will take  $U^0 = 3.0V$ and $\alpha = 1.2 V$ to approximate an OCV function for commercial lithium-ion battery systems consisting of a metal oxide positive electrode and a graphite negative electrode. With this simplification, the branch currents can be re-written as:

\begin{align}
    \label{eqn:i_a}
    I_a(t) &= +\frac{\alpha}{\mathcal{R}}\Delta z(t) + \frac{R_b}{\mathcal{R}} I(t)\\
    \label{eqn:i_b}
    I_b(t) &= \underbrace{-\frac{\alpha}{\mathcal{R}}\Delta z(t)}_{I_{\mathrm{rebalance}}(t)} + \underbrace{\frac{R_a}{\mathcal{R}} I(t)}_{I_{\mathrm{Ohmic}}(t)},
\end{align}

where $\Delta z \triangleq z_a - z_b$ and time-dependent terms have been made explicit for clarity. The first terms in each equation can be thought of as SOC rebalancing currents while the second terms are due to purely Ohmic effects (i.e. resistance mismatches). Equations (\ref{eqn:i_a}) and (\ref{eqn:i_b}) highlight the fact that, in the presence of SOC imbalance, internal currents will flow even in the absence of an externally applied current. Note that the SOC rebalancing current increases with the slope of the OCV function $\alpha$ and decreases with the total system resistance $\mathcal{R}$.

\section{Derivation of Imbalance Dynamics}
\label{sec:imbalance}

\begin{figure} 
\begin{center}
\includegraphics[width=8.4cm]{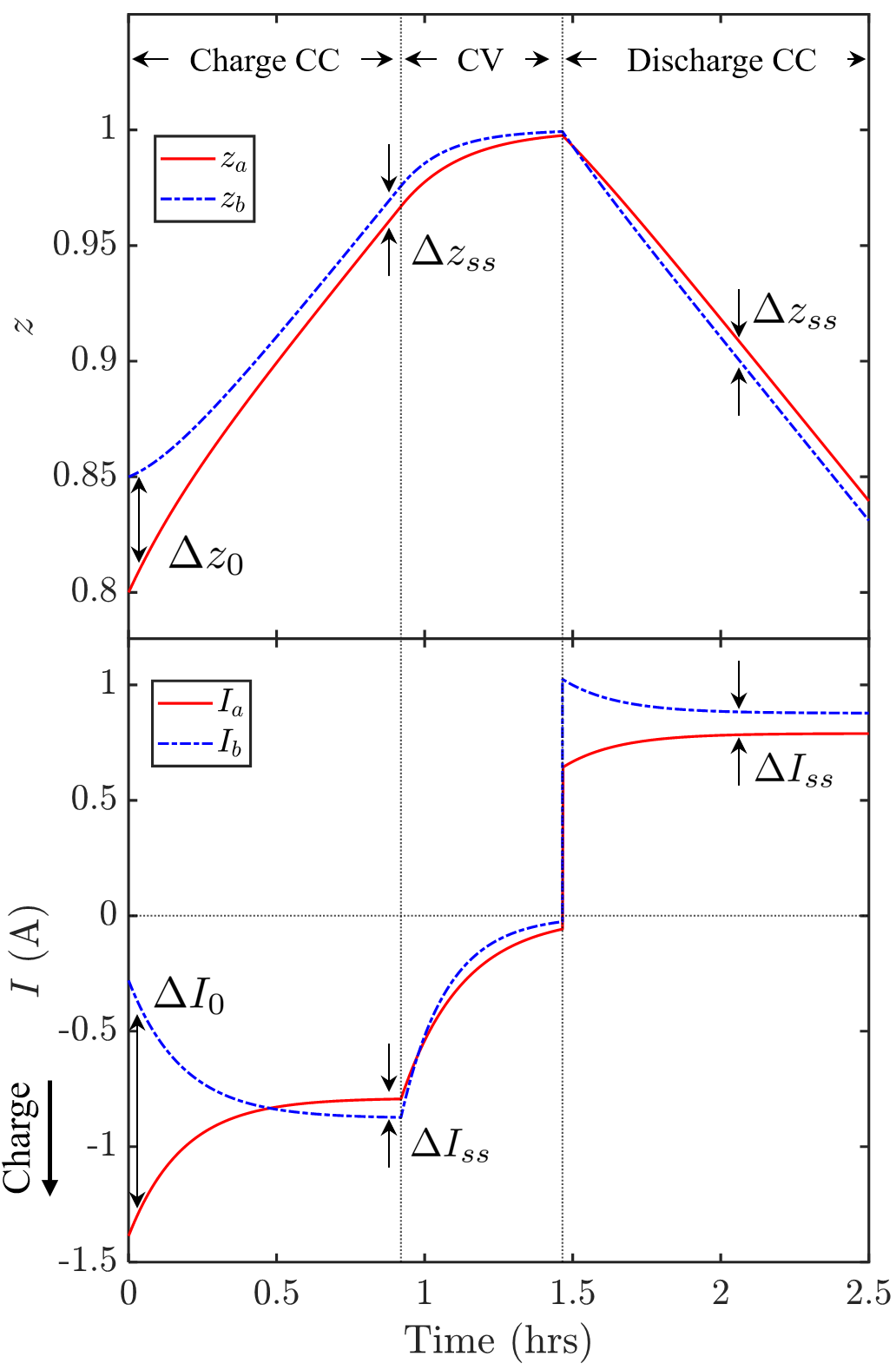} 
\caption{Example solution to the SOC and current imbalance dynamical system. The mathematical formulas used to construct this figure is provided in Section \ref{sec:imbalance} of this paper.}
\label{fig:charge_discharge}
\end{center} 
\end{figure}

In this section, we will derive analytical equations that describe the SOC and current imbalance dynamics for two parallel-connected batteries throughout a full charge-discharge cycle. The key features of the resulting system dynamics are previewed in Figure \ref{fig:charge_discharge}, which shows the imbalance dynamics for an aged battery $a$ connected in parallel with a fresh battery $b$. The mathematical formalism developed in this section will fully describe the dynamics highlighted in this figure and enable a deeper analysis of the system properties.

\subsection{SOC Imbalance}

The SOC dynamics for each battery is described by the integrator state

\begin{equation}
    \label{eqn:zdot}
    \dot{z}_i = -\frac{1}{Q_i} I_i
    .
\end{equation}

Combining (\ref{eqn:zdot}) with (\ref{eqn:i_a}) and (\ref{eqn:i_b}) yields the following system of equations:

\begin{equation}
    \label{eqn:zdot_full}
    \begin{bmatrix}
    \dot z_a \\ \dot z_b
    \end{bmatrix}
    = \frac{\alpha}{\mathcal{R}}
    \begin{bmatrix}
    \frac{-1}{Q_a} & \frac{+1}{Q_a} \\[4pt]
    \frac{+1}{Q_b} & \frac{-1}{Q_b}
    \end{bmatrix}
    \begin{bmatrix}
    z_a \\ z_b
    \end{bmatrix}
    - \frac{1}{\mathcal{R}}
    \begin{bmatrix}
    \frac{R_b}{Q_a} \\[4pt]
    \frac{R_a}{Q_b}
    \end{bmatrix}
    I
    .
\end{equation}

The SOC imbalance dynamics can then be written down by taking the difference between the two states defined in (\ref{eqn:zdot_full}) which yields

\begin{equation}
    \label{eq:dzdot}
    \Delta \dot z(t) = -\underbrace{\frac{\alpha}{\mathcal{R}}\left(\frac{1}{Q_a} + \frac{1}{Q_b}\right)}_{\text{$1/\tau$}}\Delta z(t) +
    \underbrace{\frac{1}{\mathcal{R}}\left(\frac{R_a}{Q_b} - \frac{R_b}{Q_a}\right)}_{\text{$\kappa/\tau$}} I              
    .
\end{equation}

This is a standard linear time-invariant (LTI) system which can be readily solved under a galvanostatic (constant-current) mode of operation:

\begin{equation}
    \label{eq:dz}
    \Delta z(t) = \Delta z_0 e^{-t/\tau} - \kappa(e^{-t/\tau}-1)I,
\end{equation}

where 

\begin{equation}
    \label{eq:tau}
    \tau = \frac{\mathcal{R}}{\alpha}\left(\frac{Q_aQ_b}{Q_a+Q_b}\right)    
\end{equation}

\begin{equation}
    \label{eq:kappa}
   \kappa = \frac{1}{\alpha}\left(\frac{R_aQ_a - R_bQ_b}{Q_a+Q_b}\right)
\end{equation}

and $\Delta z_0 = z_{a,0} - z_{b,0}$ is the initial SOC imbalance in the system. This system has a single negative eigenvalue $\lambda = -1/\tau$ and is therefore globally exponentially stable.

\subsubsection{Steady-State Properties}

The SOC imbalance value will decay to a steady-state value given by

\begin{align}
    \label{eq:dz_ss_1}
    \Delta z_{ss} & = \kappa I.
\end{align}

A steady-state SOC imbalance will therefore develop under an applied current even under the absences of an initial SOC imbalance. The magnitude of this imbalance will increase with the current input magnitude and decrease with the slope of the OCV function $\alpha$ due to (\ref{eq:kappa}). An interesting implication of this result is that battery chemistries with shallower OCV function slopes, such as lithium-iron phosphate (LFP) batteries, will experience more steady-state SOC imbalance during operation. However, towards the end of the charge curve, the same OCV function will typically rise sharply, accelerating the decay towards zero SOC imbalance. We also note that the steady-state SOC imbalance will approach zero as $R_aQ_a$ approaches $R_bQ_b$ due to (\ref{eq:kappa}). This scenario resembles the case of aged batteries in which resistance rise typically occurs in tandem with capacity decreases.

\subsubsection{Convergence Rate}

The convergence rate of the system towards $\Delta z_{ss}$ is understood through (\ref{eq:tau}), which predicts that SOC rebalancing times decrease when battery capacities and resistances are lower, or with steeper OCV function slopes.

A key question is whether the steady-state condition is realizable over the course of a single charge or discharge cycle. If it is, then the steady-state values may be a useful metric to consider under certain degradation modeling contexts. For example, the steady-state imbalance values could be used as a proxy for the total damage accrued over a single charge-discharge cycle. To evaluate whether or not a given steady-state imbalance value is realizable by the end of a single cycle, we consider a minimum observation time of $3\tau$, by which point the imbalance will have progressed to 95\% of its steady-state value. The C-rate must therefore be sufficiently low to enable this minimum observation time. The necessary condition for observing steady-state imbalance can be summarized as

\begin{equation}
    \label{eq:crate}
    \mathrm{C}_{\mathrm{rate}} < \frac{z_{\mathrm{max}} - z_{\mathrm{min}}}{3\tau}.
\end{equation}

In the above equation, we introduced a multiplicative factor $z_{\mathrm{max}} - z_{\mathrm{min}}$ which represents the range of SOCs accessed for this particular cycle, where $z_{\mathrm{max}}$ and $z_{\mathrm{min}}$ are the maximum and minimum SOCs accessed during the cycle. This additional term accounts for the fact that shallower SOC cycling will require even lower C-rates to observe the steady-state condition before the cycle ends. The condition described by (\ref{eq:crate}) can often be realized under practical applications. For example, consider a system with $(Q_a, R_a)$ = (4 Ah, 35 m$\Omega$) and $(Q_b, R_b)$ $<$ (5 Ah, 25 $m\Omega$). For this system, $3\tau = 0.33$ hours, and we will further assume that the system charges from 67\% to 100\% SOC so that $z_{\mathrm{max}} - z_{\mathrm{min}}$ = 0.33. Application of (\ref{eq:crate}) predicts that a C-rate of less than 1C is sufficient for the steady-state value to be realized before charging ends. By comparison, Level 2 charging for electric vehicles typically take between 3 to 8 hours, which corresponds to an equivalent C-rate well under 1C.

\subsection{Current Imbalance}
\label{sec:current}

An expression for $\Delta I \triangleq I_a - I_b$ can now be obtained by direct substitution of (\ref{eq:dz}) into (\ref{eqn:i_a}) and (\ref{eqn:i_b}) to yield

\begin{align}
    \label{eq:di}
    \Delta I(t) &= \underbrace{\frac{2\alpha}{\mathcal{R}}\Delta z(t)}_{\Delta I_{\mathrm{rebalance}}(t)} - \underbrace{\frac{\Delta R}{\mathcal{R}} I}_{\Delta I_{\mathrm{Ohmic}}(t)},
\end{align}

where $\mathcal{R} \triangleq R_a + R_b$ as before, and 

\begin{equation}
    \Delta R \triangleq R_a - R_b 
    .
\end{equation}

Equation (\ref{eq:di}) provides several useful interpretations of current imbalance. First, the current imbalance can now be readily interpreted as the output equation of the SOC imbalance state equation (\ref{eq:dz}) in a state-space representation. The current imbalance dynamics therefore inherits the stability properties, convergence properties, and time constant of the SOC imbalance system. Second, the current imbalance can be viewed as the difference between the SOC rebalancing current (first term) and the Ohmic contribution due to the resistance mismatch (second term). These two terms have opposite signs, making it possible to achieve zero steady-state current imbalance even under the presence of a non-zero SOC imbalance.

The steady-state current imbalance can be written as

\begin{equation}
    \Delta I_{ss} = \frac{2\alpha}{\mathcal{R}}\Delta z_{ss} - \frac{\Delta R}{\mathcal{R}} I.
\end{equation}

After substitution of $\Delta z_{ss}$ from (\ref{eq:dz_ss_1}), it can be shown that 

\begin{equation}
    \label{eqn:di_ss}
    \Delta I_{ss} = \left(\frac{Q_a-Q_b}{Q_a+Q_b}\right)I.
\end{equation}

Equation (\ref{eqn:di_ss}) predicts that capacity-matching is sufficient for guaranteeing zero steady-state current imbalance, irrespective of the degree of resistance mismatch. To better understand this counter-intuitive result, consider the fact that both the SOC rebalancing current and the Ohmic current increase as the resistance mismatch between the two batteries increase, but since these two terms have opposite signs, the net effect of the resistance mismatch on current imbalance is nullified. This `exact-cancellation' result is due to the assumption of affine OCV and is not generally expected to hold in the case of non-linear OCV functions (see Section \ref{sec:nonlin}).

\subsection{Potentiostatic Mode}

So far, we have focused on deriving equations that are valid under the constant-current (i.e. galvanostatic) mode of operation. However, battery charging systems typically switch to a constant-voltage (i.e. potentiostatic) mode towards the end of a charge cycle when some upper voltage limit is reached. During the constant-voltage charging stage, the total system current decays toward zero as the individual battery SOCs asymptotically approach unity. This section focuses on deriving the imbalance equations for the constant-voltage mode of operation. 

To start, we invert the input and output from (\ref{eqn:vt}) to express the total current $I$ as a function of a fixed voltage set-point $U^f$, which, when combined with the affine OCV assumption (\ref{eqn:ocv-affine}), yields

\begin{equation}
    \label{eq:current_during_cv}
    I = \frac{\alpha(z_aR_b + z_bR_a) - \Delta U\mathcal{R}}{R_aR_b}
\end{equation}

where $\mathcal{R} = R_a + R_b$ as defined before, and $\Delta U \triangleq U^f - U^0$. Substituting (\ref{eq:current_during_cv}) into (\ref{eqn:ia}) and (\ref{eqn:ib}) leads to two branch current expressions:

\begin{align}
    \label{eq:iacv}
    I_i(t) &= \frac{\alpha z_i(t) - \Delta U}{R_i}
\end{align}

where $i \in \{a,b\}$. The SOC dynamics can then be obtained by substituting (\ref{eq:iacv}) into (\ref{eqn:zdot}), which yields

\begin{equation}
    \label{eq:zcv}
    \begin{bmatrix}
    \dot z_a \\ \dot z_b
    \end{bmatrix}
    =
    \begin{bmatrix}
    -\frac{\alpha}{Q_aR_a} & 0 \\
    0 & -\frac{\alpha}{Q_bR_b}
    \end{bmatrix}
    \begin{bmatrix}
    z_a \\ z_b
    \end{bmatrix}
    +
    \begin{bmatrix}
    \frac{1}{Q_aR_a} \\ 
    \frac{1}{Q_bR_b}
    \end{bmatrix}
    \Delta U.
\end{equation}

Equation (\ref{eq:zcv}) is another LTI system which is readily solved to obtain:

\begin{align}
    \label{eq:zacv}
    z_i(t) &= z_{i,0}e^{-t/\tau_i} - \frac{1}{\alpha}(e^{-t/\tau_i} -1)\Delta U
\end{align}    

where $i \in \{a, b\}$ and the time constants $\tau_i$ are defined by

\begin{align}
    \tau_i \triangleq \frac{Q_iR_i}{\alpha}.
\end{align}

The SOC and current imbalance dynamics during potentiostatic mode of operation can then be obtained directly from (\ref{eq:iacv}) and (\ref{eq:zacv}).

\subsection{Summary of the Imbalance System Behavior}

We now have the means to quantitatively understand the system characteristics shown in Figure \ref{fig:charge_discharge}. In this scenario, the battery parameters used were $(Q_a, R_a)$ = (5 Ah, 50 m$\Omega$), and $(Q_b, R_b)$ = (5.6 Ah, 33 m$\Omega$). The current input was $I=-1.67$A during charge and $I=1.67A$ during discharge, the maximum voltage was set to 4.2V, and the constant-voltage hold cut-off condition was set to 83 mA. Battery $a$ started at 5\% lower SOC than battery $b$. The system then underwent a constant-current, constant-voltage charging protocol, followed by a constant-current discharge. The persistent SOC and current imbalance observed during both charge and discharge can now be quantitatively explained by (\ref{eq:dz_ss_1}) and (\ref{eqn:di_ss}). The initially large current imbalance at the beginning of the charge cycle is due to the SOC rebalancing current which forced current from battery $b$ to flow into battery $a$. Toward the end of the charge cycle, the total current imbalance can be shown to include contributions from both a steady-state SOC rebalancing current and a steady-state Ohmic current. The SOC imbalance decreased during the constant-voltage hold step, leading to a smaller spike in current imbalance at the beginning of the discharge step.

\section{Discussion}

This section dives deeper into several aspects of the current and SOC imbalance system properties. Section \ref{sec:qr} introduces the concept of $(q,r)$ maps to answer questions about which battery will see higher currents and SOCs. Section \ref{sec:nonlin} comments on the validity of the affine OCV assumption. Finally, Section \ref{sec:aging} briefly discusses the implications of SOC and current imbalance on long-term aging behavior of parallel-connected battery systems.

\subsection{$(q,r)$ Maps: Which Battery Will Experience Higher Currents and SOCs?}
\label{sec:qr}

\begin{figure} 
    \begin{center}
    \includegraphics[width=7.4cm]{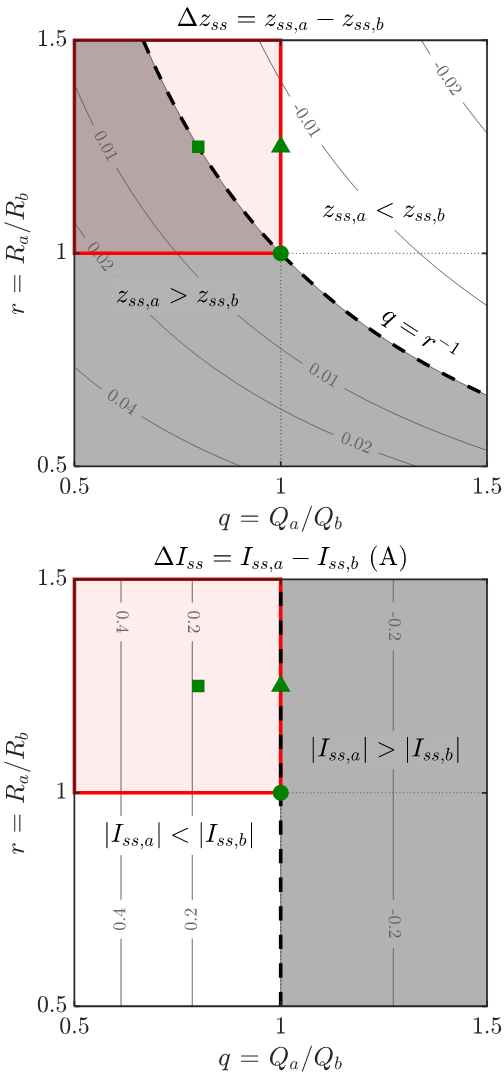} 
    \caption{Sensitivity of steady-state SOC imbalance ($\Delta z_{ss}$) and current imbalance ($\Delta I_{ss}$) to mismatches in battery capacities and resistances during charging. The top left quadrant of each plot (red boxes) represents an aged battery $a$ connected in parallel to a fresh battery $b$, with $Q_a<Q_b$ and $R_a>R_b$. Solutions are given for $I=-1.67$ A (C/3) and with $(Q_a,R_a)$ = (5 Ah, 50m$\Omega$).} 
    \label{fig:steady_state}
    \end{center} 
\end{figure}

\begin{figure*} 
    \begin{center}
    \includegraphics[width=16.8cm]{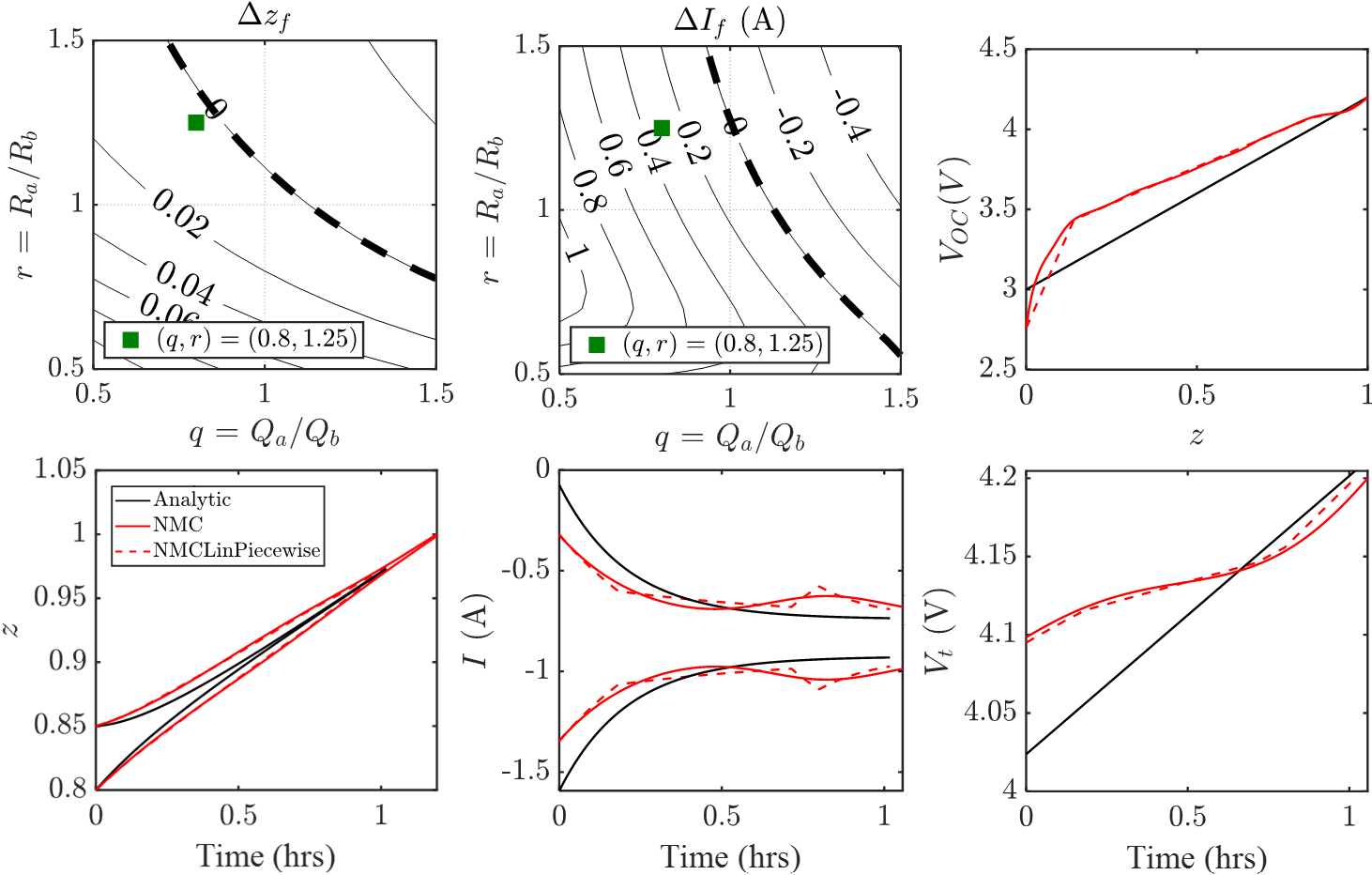} 
    \caption{SOC and current imbalance dynamics for an NMC/graphite system with a non-linear OCV function. $\Delta z_f$ and $\Delta I_f$ are the final SOC and current imbalance values after the terminal voltage reaches 4.2V on charge. The timeseries plots show the evolution of SOC and current over a single charge cycle, comparing between simulation results using affine OCV functions (black), non-linear OCV functions (solid red), and piecewise-affine OCV functions (dashed red). All simulations use $I=-1.67$A (C/3), $z_{a,0} = 0.85$, $z_{b,0} = 0.90$, $Q_a = $ 5 Ah, $R_a = 50$ m$\Omega$, and $(q,r) = (0.8, 1.25)$.}
    \label{fig:nonlinear_study}
    \end{center} 
\end{figure*}

Here, we will use Figure \ref{fig:steady_state} to understand how different combinations of resistances and capacities affect the steady-state behavior of the imbalance system. Defining 

\begin{align}
\label{eq:rq}
r &\triangleq R_a/R_b \qquad\qquad q \triangleq Q_a/Q_b ,
\end{align}

a map of $\Delta z_{ss}$ and $\Delta I_{ss}$ as a function of $q$ and $r$ (\ref{eq:dz_ss_1},\ref{eqn:di_ss}) can then be built. The figure highlights regions in ($q,r$)-space under which imbalance values are either positive or negative. Each subplot is centered at $(1,1)$ which corresponds to a system having two perfectly-matching batteries (circle marker). The top left quadrants (red boxes) correspond to regions in which battery $a$ is more aged than battery $b$, with $Q_a < Q_b$ and $R_a > R_b$. The contour labels indicate the amplitude of the steady-state values during charge.

The top panel shows that zero steady-state SOC imbalance is achieved under the condition $q = r^{-1}$, or equivalently, $R_aQ_a = R_bQ_b$ (\ref{eq:dz_ss_1}). Hence, zero steady-state SOC imbalance is achievable even in systems with mismatched capacities and resistances. In particular, as battery $a$ ages, $R_a$ increases while $Q_a$ decreases, so $R_aQ_a$ may remain sufficiently close to $R_bQ_b$ such that the steady-state SOC imbalance remains close to zero. A system described by $(0.8, 1.25)$ (square marker), for example, represents an aged battery with 20\% capacity loss and 25\% resistance growth which achieves exactly zero steady-state SOC imbalance. In general, the condition $q > r^{-1}$ must be satisfied in order for the aged battery $a$ to experience a comparatively lower SOC at the end of a charge cycle.

The bottom panel shows that $q = 1$ is sufficient for attaining zero steady-state current imbalance and irrespective of the value of $r$ (see Section \ref{sec:current}). This result implies that an aged battery $a$, when connected with a fresh battery $b$ (red region), will always experience a lower steady-state current magnitude for any $r$ provided that $q<1$.

The $(q,r)$-map also illustrates the existence of scenarios in which steady-state SOC imbalance persists even in the absence of current imbalance, and vice versa. For example, the system $(0.8, 1.25)$ (square markers) achieves zero SOC imbalance, yet still experiences a current imbalance of 0.2A at steady-state. Meanwhile, the system $(1, 1.25)$ (triangle markers) reaches zero current imbalance, yet, an SOC imbalance persists at steady-state. We finally note that the numerical values in Fig. \ref{fig:steady_state} are provided for the case of charging. During discharge, values in the $(q,r)$-maps will flip signs. A similar analysis can then be repeated to show that an aged battery $a$, when connected to a fresh battery $b$, will continue to experience lower steady-state current magnitudes for any $r$ and $q<1$, and that battery $a$ will also experience higher steady-state SOCs (i.e. be less discharged) when $q > r^{-1}$.

\subsection{Comments on the Affine OCV Assumption}
\label{sec:nonlin}

A key assumption used in our derivation was that the battery OCV functions were affine and identical. Here, we explore the applicability of the linear model towards more practical battery systems having non-linear OCV curves. To begin, we numerically simulated the imbalance dynamics for two additional cases: a non-linear OCV function and a piecewise-affine OCV function. The non-linear OCV function was adapted from \cite{Chen2020d} and represents a battery comprising a nickel manganese cobalt (NMC) positive electrode and graphite negative electrode. The piecewise-affine OCV function consisted of four linear segments which approximate the non-linear OCV function. The individual battery SOC and current dynamics were then solved numerically in discrete time using using a forward Euler approach.

The simulation results are shown in Figure \ref{fig:nonlinear_study}. The first row shows the $(q,r)$-maps for SOC imbalance ($\Delta z_f$) and current imbalance ($\Delta I_f$). These values correspond to the final imbalance value when the maximum voltage of 4.2V was reached and are analogous to the steady-state values $\Delta z_{ss}$ and $\Delta I_{ss}$ reported previously. However, in this case, steady-state convergence is not guaranteed due to the non-linear OCV function. We observe that the non-linear system behaves in a qualitatively similar way compared to the non-linear system, but the numerical values differ. Now, the system $(0.8, 1.25)$ no longer achieves zero `steady-state' SOC imbalance as was the case for the linear system, and the `steady-state' current imbalance has become dependent on $r$. To further study these differences, we plotted the evolution of the SOC and current imbalance over time for the case $(q,r)=(0.8,1.25)$. The analytic (linear) result underestimated the degree of SOC imbalance and could not capture the inflection point in the current imbalance towards the end of charge. The system using a piecewise-affine function, however, was successful at approximating the non-linear case. 

Overall, we observe that the model assuming an affine OCV function can provide correct qualitative trends compared to a system with a non-linear OCV function. The numerical accuracy of the linear model will, however, expectedly deteriorate for systems having large non-linearities in the OCV function.

\subsection{From Current and SOC Imbalance to Divergent Aging}
\label{sec:aging}

SOC and current imbalances can act as precursors to possible divergent aging over cycle life. To motivate further work in this area, we will briefly highlight several aging scenarios to illustrate the rich connection between imbalance dynamics and degradation dynamics. Recall that, in Figure \ref{fig:steady_state}, we identified the fact that an aged battery $a$ will always experience lower current magnitudes at steady-state compared to the fresh battery $b$. The lower current magnitude causes battery $a$ to more protected against lithium plating compared to battery $b$ over life, since the negative electrode overpotential for battery $a$ will be comparatively higher at the end of charge \citep{Yang2017}. Now, consider the possibility that $q<r^{-1}$, in which case the aged battery $a$ will additionally realize higher SOCs at the end of charge compared to battery $b$. The positive electrode of battery $a$ will therefore be more delithiated at the end of charge compared to battery $b$. A more delithiated positive electrode is associated with strain-induced loss of active material \citep{O_Kane_2022}, so battery $a$ will experience more active material loss compared to battery $b$. Hence, we have identified a scenario under which the aged battery $a$ is more protected against lithium plating, while simultaneously being less protected against active material loss.  The overall degradation rate of battery $a$ will depend on the degradation modeling assumptions, such as the relative contribution of lithium plating versus active material losses to the total capacity loss \citep{O_Kane_2022}. In summary, imbalance dynamics will inevitably manifest as differences in degradation trajectories. The analytic description of the imbalance dynamics established in this paper sets the stage for more detailed studies to answer the ultimate question: ``under what conditions do aging trajectories converge in parallel-connected battery systems?''

\section{Conclusion and Future Work}

In this work, we derived state equations for SOC and current imbalance dynamics in two parallel-connected batteries assuming affine and identical OCV functions (\ref{eq:dz},\ref{eq:di}). The steady-state values for SOC and current imbalance (\ref{eq:dz_ss_1},\ref{eqn:di_ss}) reveal the conditions under which an aged battery will experience higher SOCs and more currents at the end of a charge cycle. Importantly, we found that an aged battery, when connected in parallel with a fresh battery, will always experience lower current magnitudes at the end of a charge or discharge cycle. We relaxed the assumption of an affine OCV function to show that the steady-state analysis can qualitatively capture the correct trends when compared against the general non-linear case. We finally discussed how the steady-state SOC and current imbalance values can be thought of as precursors to heterogeneous aging in parallel-connected packs. 

This work represents a necessary step towards degradation modeling in parallel-connected battery systems. From here, battery aging dynamics can be further developed, which leverages the analytical equations derived in this paper to provide a cycle-by-cycle accounting of how capacities and resistances of the individual batteries evolve, e.g. due to differences in steady-state SOC and current imbalance. The derivation described in this work can also be generalized to the case of an arbitrary number of cells connected in parallel. Finally, model parameters such as resistance can be augmented with temperature and SOC-dependence to improve model accuracy under real-world scenarios.

\bibliography{main} 

\end{document}